\documentclass[twocolumn,12pt]{article}
\usepackage{graphics}

\date{}
\def\onehead#1{\vskip\baselineskip\centerline{\normalsize #1}
               \vskip8pt}

\title{\vspace*{-15mm}
{\small XXVI International Conference on High Energy Physics, Dallas, TX
(6--12 August 1992) 1058--1061}\\
\vskip 12pt
{\normalsize FERMILAB E791}
   \thanks{This work was supported by the U.S. D.O.E. and N.S.F.,
   the U.S.-Israel Binational Science Foundation, and the Brasilian
   Conselho Nacional de Desenvolvimento Cient\'\i fico e Tecnol\'{o}gico.}}

\author{\null \vspace*{-35pt} \\
{\normalsize{L.M.~Cremaldi,$^6$}
E.M. Aitala,$^6$
F.M.L. Almeida,$^9$
S.\ Amato,$^1$
J.C. Anjos,$^1$
J.A.~Appel,$^4$}\\ 
\null \vspace*{-21pt} \\
{\normalsize{D.~Ashery,$^{10}$}
J.~Astorga,$^{12}$
S.~Banerjee,$^4$
S.~Beck,$^{10}$
I.~Bediaga,$^1$
G.~Blaylock,$^2$
S.B.~Bracker,$^{11}$}\\ 
\null \vspace*{-21pt} \\
{\normalsize{P.R.~Burchat,$^2$}
R.~Burnstein,$^5$
T.~Carter,$^4$
I.~Costa,$^1$
K.~Denisenko,$^4$
C.~Darling,$^{14}$
P.~Gagnon,$^2$}\\
\null \vspace*{-21pt} \\
{\normalsize{S.~Gerzon,$^{10}$}
K.~Gounder,$^6$
D.~Granite,$^7$
M.~Halling,$^4$
C.~James,$^4$
P.A.~Kasper,$^5$
S.~Kwan,$^4$}\\
\null \vspace*{-21pt} \\
{\normalsize{J.~Lichtenstadt,$^{10}$}
B.~Lundberg,$^4$
J.~R.~T.~de~Mello~Neto,$^1$
R.~Milburn,$^{12}$
J.M.\ de Miranda,$^1$}\\
\null \vspace*{-21pt} \\
{\normalsize{A.~Napier,$^{12}$}
A.~Nguyen,$^7$
A.B.d'Oliveira,$^3$
K.C.~Peng,$^5$
M.V.~Purohit,$^8$
B.Quinn,$^6$
S.~Radeztsky,$^{13}$}\\
\null \vspace*{-21pt} \\
{\normalsize{A.~Rafatian,$^6$}
A.J.~Ramalho,$^9$
N.W.~Reay,$^7$
K.~Reibel,$^7$
J.J.~Reidy,$^6$
H.~Rubin,$^5$
A.~Santha,$^3$}\\
\null \vspace*{-21pt} \\
{\normalsize{A.F.S.~Santoro,$^1$}
A.~Schwartz,$^8$
M.~Sheaff,$^{13}$
R.A.~Sidwell,$^7$
H.~daSilva~Carvalho,$^9$
J.~Slaughter,$^{14}$}\\
\null \vspace*{-21pt} \\
{\normalsize{M.D.~Sokoloff,$^3$}
M.~Souza,$^1$
N.~Stanton,$^7$
K.~Sugano,$^2$
D.J. Summers,$^6$
S.\ Takach,$^{14}$
K.\ Thorne,$^4$}\\
\null \vspace*{-21pt} \\
{\normalsize{A.~Tripathi,$^7$}
D.~Trumer,$^{10}$
S.~Watanabe,$^{13}$
J.~Wiener,$^8$
N.\ Witchey,$^7$
E.\ Wolin,$^{14}$
D.\ Yi$^6$}\\
{\normalsize{$^1$Centro Brasileiro de Pesquisas Fisicas, Rio de Janeiro}}\\
\null \vspace*{-22pt} \\
{\normalsize{$^2$University of California, Santa Cruz, CA 95064}}\\
\null \vspace*{-22pt} \\
{\normalsize{$^3$University of Cincinnati, Cincinnati, OH 45221}}\\
\null \vspace*{-22pt} \\
{\normalsize{$^4$Fermilab, Batavia, IL 60510}}\\
\null \vspace*{-22pt} \\
{\normalsize{$^5$Illinois Institute of Technology, Chicago, IL 60616}}\\
\null \vspace*{-22pt} \\
{\normalsize{$^6$University of Mississippi, Oxford, MS 38677}}\\
\null \vspace*{-22pt} \\
{\normalsize{$^7$Ohio State University, Columbus, OH 43210}}\\
\null \vspace*{-22pt} \\
{\normalsize{$^8$Princeton University, Princeton, NJ 08544}}\\
\null \vspace*{-22pt} \\
{\normalsize{$^9$Universidade Federal do Rio de Janeiro}}\\
\null \vspace*{-22pt} \\
{\normalsize{$^{10}$Tel--Aviv University, Tel--Aviv 69978}}\\
\null \vspace*{-22pt} \\
{\normalsize{$^{11}$317 Belsize Drive, Toronto, Ontario M4S1M7}}\\
\null \vspace*{-22pt} \\
{\normalsize{$^{12}$Tufts University, Medford, MA 02155}}\\
\null \vspace*{-22pt} \\
{\normalsize{$^{13}$University of Wisconsin, Madison, WI 53706}}\\
\null \vspace*{-22pt} \\
{\normalsize{$^{14}$Yale University, New Haven, CT 06511}}\\
\null \vspace*{-4pt} \\
{\normalsize{Abstract}}\\
\null \vspace*{-16pt} \\
\parbox{5.5in}{\normalsize
Fermilab E791, a very high statistics charm particle experiment,
recently
completed its data taking at Fermilab's Tagged Photon Laboratory.  Over 20
billion events were recorded through a loose transverse energy trigger and
written to 8mm tape in the the 1991-92 fixed target run at Fermilab.  This
unprecedented data sample containing charm is being analysed on many-thousand
MIP RISC computing farms set up at sites in the collaboration.  A glimpse of
the data taking and analysis effort is presented.  We also show some 
preliminary
results for common charm decay modes.  Our present analysis indicates a very
rich yield of over 200K reconstructed charm decays.} \\ 
}

\begin{document}

\maketitle




\clearpage

\onehead{INTRODUCTION}

E791 is the fourth in a series of charm particle experiments performed at
Fermilab's Tagged Photon Lab (TPL) over the past several years.  The charm
sample is produced through 500 GeV/c $\pi^-N$ interactions in a 
platinum-diamond
target. The data is recorded with a low bias transverse energy trigger $(E_T)$
formed in the hadron and electro\-magnetic calorimeters.  The goal is to
reconstruct over 100K charm decays for high statistics studies and a for a
close look at rare charm   
decay physics.~As in the past,~E791 uses a high
precision silicon vertex detector 
and a large open geometry spectrometer
to extract charm decays on low backgrounds.

The success of the E791 hinges on its high data rate capability and offline
reconstruction of data.  A high rate data acquistion system was built$^1$ to 
record
a multi-billion event data sample necessary for extracting charm with such high
statistics. 
The challenge ahead is the offline reconstruction of this very
large event sample.  Computing costs are dropping at a rate of about a 
factor of
two each year, facilitating the timely processing of this large data sample.

\onehead{SPECTROMETER}

The spectrometer features 23 planes of silicon microstrip detectors covering a
$\pm$ 100 mrad solid angle in the forward direction.  The magnetic spectrometer
consists of two horizontally bending dipoles with a combined $p_T$ kick of 
about
.5 GeV/c. 35 planes of drift chambers are interleaved throughout for charged
particle tracking.  Typical momentum resolutions for high mass charm states
are in the 8-12 $MeV/c^2$ range depending on the decay multiplicity and decay
energy available.

The target consisted of a series of foils, first .5 mm Pt foil followed by four
1.5 mm C foils, each spaced 1.5 cm apart.  This arrangement optimizes the
detection of short lived states decaying in the air gaps between foils and
provides sufficient interaction rate for the experiment.  The target represents
about .03 interaction length and produces about a 40 KHz interaction rate
during our beam spill.

The beam particles are tracked into the \break 
target~by~a set of 8 upstream MWPC's
po\-sitioned in the beamline and 6 planes of silicon microstrips (SMDs) located
just upstream of the target assembly.  The beam tracking allows a precise
transverse beam location (7$\mu m)$, while the z position of the interaction 
point
can be easily isolated in a target foil.  We use this beam constraint in
locating the primary vertex interaction point with high efficiency.

Particle identification is provided by two multi-cell threshold Cherenkov
counters$^2$, giving $\pi /K$ separation in the 6-72 GeV/c momentum range.  An
electromagnetic and hadron calorimeter$^{3,4}$ provide good e/$\pi$ separation,
as
well as photon identification.  Muons are tagged in a pair of hodoscopes
following steel absorber at the downstream end of the spectrometer.

\onehead{DATA SAMPLE}

E791 recorded data with a loose transverse energy trigger $(E_T)$ and 
beam track
requirement.  The $E_T$ triggered event rate was about 9000 events/sec.  E791's
high rate data acquisition system, capable of logging 10 MB/sec., recorded
about 4000 events each second during spill and interspill.  These events were
written to 8mm tape.  24000 data tapes were recorded in the 6 month running
period, corresponding to 20 billion events.

\onehead{COMPUTING}

With such a large data sample, computing becomes a major issue.  We have so
far implemented two large RISC based computing farms at the University of
Mississippi and 
Ohio~State~University for reconstruction of E791 data.  Each
of these farms is equivalent to about 900 and 1500 MIPs respectively, with
plans for future expansions.

These startup farms went into action in February 1992, just a month after the
fixed target run had ended.  At that time, using a preliminary version of our
reconstruction code, we extracted our first charm signals.$^{5,6}$

Future farm activity is planned at Fermilab on the large IBM and Silicon
Graphics systems set up at their computing center.  In addition, the CBPF group
is adapting a set of ACPII processors at FNAL for E791 use, in a joint
FNAL-CBPF effort.  In all E791 will have over 7000 MIPs of dedicated computing
contributing to the 1-2 year reconstruction effort.

\vskip 9pt
\onehead{CHARM PHYSICS}

The physics potential from E791 is enormous.  It is the highest statistics
charm experiment done to date.  We will be able to improve the lifetime
measurements of $D$-mesons to an unprecedented accuracy and make substantial
improvements to charm baryon lifetime measurements.  Our high statistics
$D$-meson samples can be used to search for $D^0$--$\overline{D}\,{^0}$
mixing, singly and
doubly Cabibbo suppressed decays as $D^+ \rightarrow  K^+\pi^+\pi^-$,
 and $D^0 \rightarrow   K^+\pi^-.$  We will be
able to make major contributions to charm semileptonic decays, especially in
the Cabibbo suppressed decays as $D^0 \rightarrow \pi^-l^+\nu$.  Searches 
for flavor changing
neutral currents, $D^+ \rightarrow \pi^+\mu^+\mu^-,$ and for CP violation in
$D$ decays, such as
in $D^0 \rightarrow K^+ K^-$, may be made.  E791's good efficiency for 
detecting
$\Lambda,\Xi$, and
$K_s$ vee decays will give us high sensitivity to many rare charm baryon decay
channels.

Thus far our efforts have been focused on optimization of reconstruction code
and surveys of the such charm decay modes as $D\rightarrow K\pi,K2\pi$, and
$K3\pi$ decays.
For these studies, a vertex separation cut corresponding to 6-8 $\sigma_z$
is made
between primary and secondary charm vertex.  We require that the charm decay
points back to the primary or there be transverse momentum balance about the
D-meson line of flight.  Charm signals in these decay modes look very
promising and are displayed in Figure 1.  We presently estimate a reconstructed
charm yield of over 200K events into these modes.

\onehead{CONCLUSION}

E791 has reached a milestone in recording a data set with more than 200K fully
reconstructable charm decays.  We are improving our code and will begin a
physics pass on the data soon.  We are looking forward to the start of our
charm physics analyses.

\onehead{REFERENCES}

\newcounter{bean}
\begin{list}
{\arabic{bean}.}{\usecounter{bean} \setlength{\leftmargin}{6.0mm}
}

\item S. Amato et al., "The E791 Parallel Architecture Data 
Acquisition System", 
Nucl.~Inst.~Meth.~{\bf A324} (1993) 535.

\item  D. Bartlett et al., \, Nucl. Inst. Meth. {\bf A260}~(1987) 55.

\bibitem{three} J. A. Appel et al., \, Nucl. Inst. Meth. {\bf A243}~(1986) 361.

\item 
V.K.Bharadwaj et al., Nucl.~Inst.~Meth. {\bf A228} (1985) 283., D.J.~Summers, 
Nucl.~Inst. Meth. {\bf A228} (1985) 290.

\item
D.J.~Summers et al., "Charm Physics~at Fermilab E791", UMS/HEP/92-020,
May 28,~1982.~XXVII Rencontre de Moriond, Electroweak Interactions 
and
Unified The\-ories, Les Arcs, France, 15-22 March 1992.

\item
K.~Thorne et al., "Update on Hadropro- \break 
duced Charm at TPL", 
\hbox{FERMILAB-Conf-92/174.}  XXVII 
Rencontre de Moriond,
QCD~and~High~Energy Hadronic
Interactions, Les Arcs, France, March 1992.

\end{list}

\begin{figure*}
\vspace*{-2cm}
\leftline{\hspace{8mm}
\resizebox{6.06in}{!}{\includegraphics{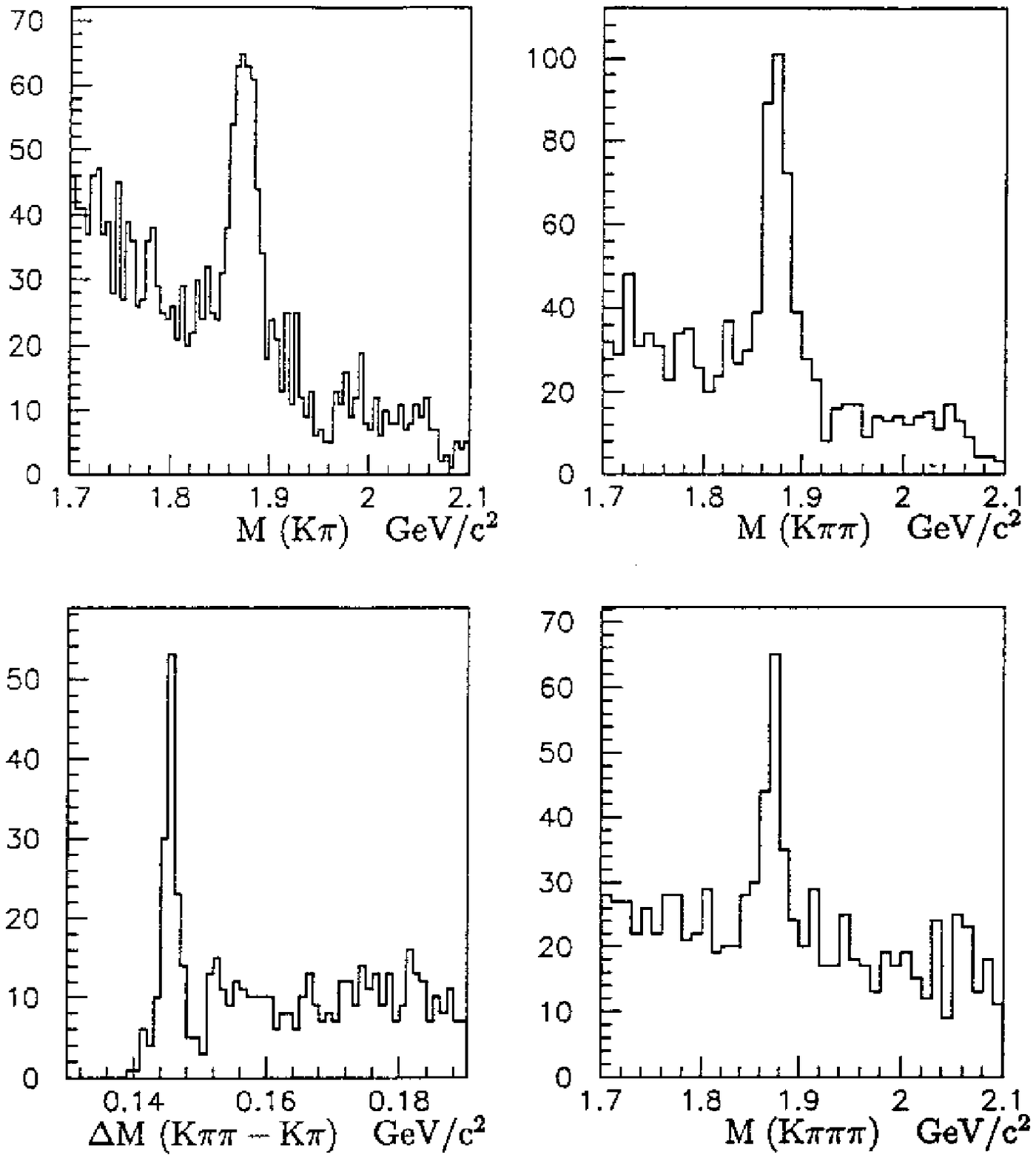}}}
\centerline{\small Figure 1. Charm signals
 extracted from a small fraction of the E791 data set.}
\end{figure*}

\end{document}